\documentclass[sigconf]{acmart}
\usepackage[linesnumbered, ruled, vlined]{algorithm2e}
\usepackage{pdfpages}
\usepackage{array}
\usepackage{graphicx}
\usepackage{stfloats}
\usepackage{hyperref}
\let\svthefootnote\thefootnote
\newcommand\freefootnote[1]{%
  \let\thefootnote\relax%
  \footnotetext{#1}%
  \let\thefootnote\svthefootnote%
}
\newif\ifshownotes
\shownotestrue %

\ifdefined\MakeWithNotes
\shownotestrue
\fi
\ifdefined\MakeWithoutNotes
\shownotesfalse
\fi

\ifshownotes
\newcommand{\colornote}[3]{{\color{#1}\bf{#2#3}\normalfont}}
\newcommand{\colornoteTwo}[3]{{\color{#1}\bf{#3}\normalfont}}
\newcommand{\colornoteThree}[2]{{\color{#1}\bf{#2}\normalfont}}      
\else
\newcommand{\colornote}[3]{}
\newcommand{\colornoteTwo}[3]{}
\newcommand{\colornoteThree}[2]{}      
\fi

\definecolor{darkgreen}{rgb}{0.0,0.75,0.0}

\newcommand{\vishnu}[1]{{#1}}

\newcommand{\change}[1]{#1}

\newcommand{\blocking}[0]{blocking }
\newcommand{\detail}[0]{detail }

\AtBeginDocument{%
  \providecommand\BibTeX{{%
    \normalfont B\kern-0.5em{\scshape i\kern-0.25em b}\kern-0.8em\TeX}}}

\setcopyright{acmlicensed}
\copyrightyear{2024}
\acmYear{2024}
\acmDOI{3654777.3676444}

\def\hyperref#1#2#3{}
\acmSubmissionID{3405}

\begin{document}

\title{Block and Detail: Scaffolding Sketch-to-Image Generation}

\author{Vishnu Sarukkai}
\affiliation{%
  \institution{Stanford University}
  \city{Stanford}
  \state{CA}
  \country{USA}
  \postcode{94305}
}

\author{Sylvia Yuan*}
\affiliation{%
  \institution{Stanford University}
  \city{Stanford}
  \state{CA}
  \country{USA}
  \postcode{94305}
}

\author{Mia Tang*}
\affiliation{%
  \institution{Stanford University}
  \city{Stanford}
  \state{CA}
  \country{USA}
  \postcode{94305}
}

\author{Maneesh Agrawala}
\affiliation{%
  \institution{Stanford University}
  \city{Stanford}
  \state{CA}
  \country{USA}
  \postcode{94305}
}

\author{Kayvon Fatahalian}
\affiliation{%
  \institution{Stanford University}
  \city{Stanford}
  \state{CA}
  \country{USA}
  \postcode{94305}
}

\renewcommand{\shortauthors}{Sarukkai, et al.}

\begin{abstract}
We introduce a novel sketch-to-image tool that aligns with the iterative refinement process of artists. 
Our tool lets users sketch {\em blocking} strokes to coarsely represent the placement and form of objects and {\em detail} strokes to refine their shape and silhouettes. 
We develop a two-pass algorithm for generating high-fidelity images from such sketches at any point in the iterative process. 
In the first pass we use a ControlNet to generate an image that strictly follows all the strokes (blocking and detail) and in the second pass we add variation by renoising regions surrounding blocking strokes.
We also present a dataset generation scheme that, when used to train a ControlNet architecture, allows regions that do not contain strokes to be interpreted as not-yet-specified regions rather than empty space. 
We show that this partial-sketch-aware ControlNet can generate coherent elements from partial sketches that only contain a small number of strokes.
The high-fidelity images produced by our approach serve as scaffolds that can help the user adjust the shape and proportions of objects or add additional elements to the composition.
We demonstrate the effectiveness of our approach with a variety of examples and evaluative comparisons. 
\vishnu{
Quantitatively, evaluative user feedback indicates that novice viewers prefer the quality of images from our algorithm over a baseline Scribble ControlNet for 84\% of the pairs and found our images had less distortion in 81\% of the pairs. 
}

\end{abstract}

\begin{teaserfigure}
\centering
\vspace{-1em}
\includegraphics[width=\textwidth]{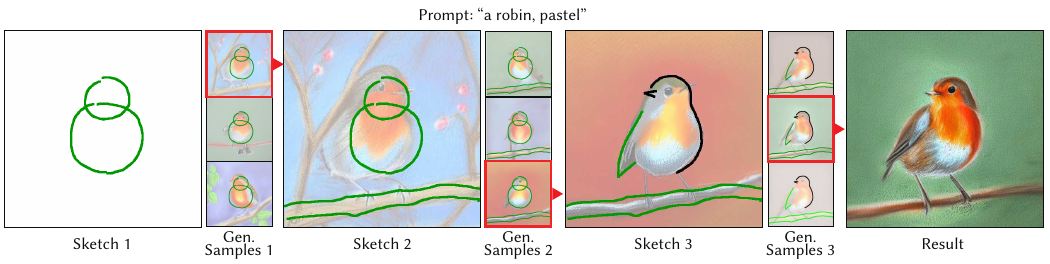}
\vspace{-2em}
\caption{
Our sketch-to-image generation tool supports the iterative refinement process artists use to create images. Users can sketch blocking strokes (green) to specify coarse spatial composition and detail strokes (black) to specify more precise silhouettes and shapes. 
At any point, users can generate high-fidelity image samples that loosely follow the spatial structure of blocking strokes portraying variations in the shape and proportions of the objects the user has blocked out (e.g. the position and size of the robin after sketch~1, the placement of the branch after sketch~2.)  The generated samples tightly match the spatial structure of detail strokes preserving their shape as much as possible (e.g. drawing the right contour of the robin and its beak specifies the pose and silhouette of the bird after sketch~3). After each sketch, the generated image samples serve as scaffolding for the user to decide how they might update their sketch by adding or adjusting strokes. For example, the generated samples after sketch~2 (gen. samples 2) present the user with varying head orientations and body silhouettes. The user chooses to overlay the bottom image from generative samples 2 over the canvas, and uses it as a guide to add detail strokes in sketch~3. 
}
\label{fig:robin}
\end{teaserfigure}

\maketitle

\freefootnote{* indicates equal contribution.}

\section{Introduction}

Artists create drawings through a process of iterative refinement. 
They commonly begin by \emph{blocking} out key pieces of the overall composition, often sketching simple 2D primitives (e.g. circles, polygons) or contours to roughly represent the placement and form of objects. Then they add \emph{detail} by drawing strokes that refine the silhouette of objects or depict subparts and interior regions~\cite{novick:1987:cogconstraints, liu:2024:dynamicsketch,dodson1990keys,barrett2021}.

Importantly, each new stroke, whether it is intended for blocking or detail, is more than just an addition to the drawing, \emph{it is a source of critical feedback to the artist}~\cite{hertzmann:2022:modeling,tversky:suwa:2009:thinking}. Artists constantly inspect the partial state of their drawing and use their reaction to inform decisions about where to place new strokes or how to adjust prior ones. This process of interacting with a partial drawing, which involves constant observation of current results along with interleaved phases of sketching for blocking and detail, is fundamental to an artist ultimately arriving at their preferred depiction of a subject.

Recent sketch-to-image tools\,\cite{voynov:2023:sketchdiff,zhang:2023:controlnet,balaji:2023:ediffi} offer an exciting new approach to image generation as a user only needs to provide a sketch, and the tool can generate high-fidelity detailed imagery in a variety of styles (e.g. photographic, watercolor, illustration) that closely resembles the provided sketch. However, these tools do not fully support the interactive process of working with partial sketches that artists 
commonly use, and thereby miss opportunities to assist the process of creating imagery.

\begin{figure}
\centering
\includegraphics[width=3.3in]{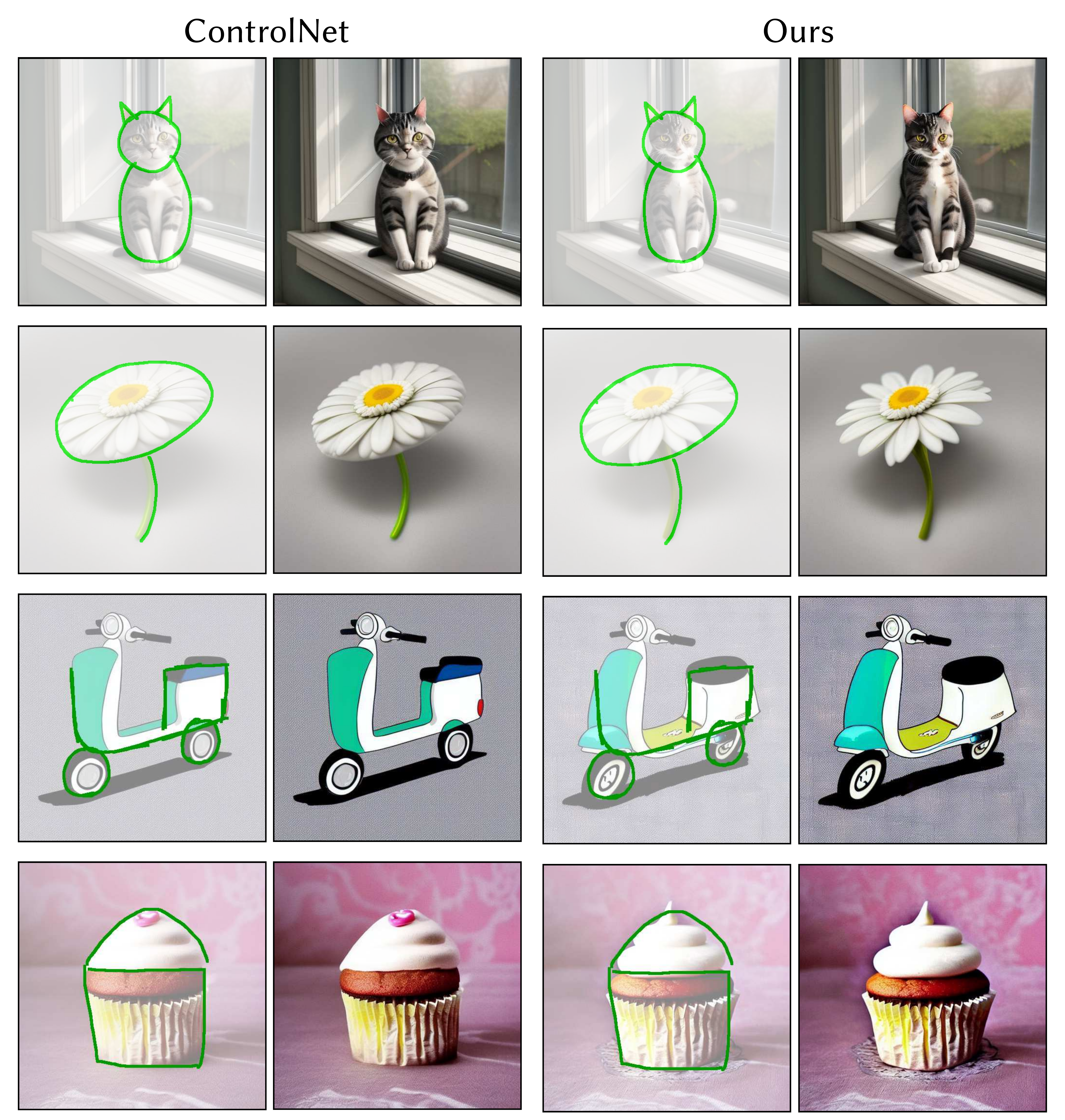}
\caption{
\vishnu{In early stages of sketching, artists often specify object forms via rough blocking strokes. Standard ControlNet adheres too strictly to these strokes, creating images with object forms that are unrealistic or poorly proportioned (misshaped cat, overly circular flower, poorly proportioned scooter, simplified cupcake silhouette). Instead, our algorithm treats blocking strokes as rough guidelines for object form, enabling artists to generate visual inspiration that is both realistic and accurately matches their intent. }
}
\label{fig:blockingfig}
\end{figure}

First, existing methods often fail to generate output that aids exploration of proportion and form typical of blocking. Prior work either forces generated images to conform too tightly to the sketch (they treat all strokes as detail strokes) or they conform too little, allowing the output to significantly diverge from blocking strokes (e.g., with an amount governed by unpredictable knobs like ``guidance strength''). 
\vishnu{In the former case, strict adherence to blocking strokes yields high-fidelity output that echoes the oversimplified or unsatisfactory forms the user is working to improve. For example, given blocking strokes, ControlNet~\cite{zhang:2023:controlnet} generates images with unrealistic or poorly proportioned forms Fig~\ref{fig:blockingfig}-ControlNet). Instead we seek a system that loosely conforms to blocking strokes, offering alternative forms that are more plausible and realistic (Fig~\ref{fig:blockingfig}-Ours).}
In the latter case of too little conformance, the system frustratingly disregards key details of forms the user has put time into creating.

Second, existing methods do not anticipate \emph{partial sketches} as input. They treat regions without strokes as empty regions of a composition, not as regions that are yet to be drawn. As a result, generated output often fails to provide examples of how to more completely specify an in-process object or depict other elements might be added to a composition. 

In this paper we present a sketch-to-image tool that is designed to scaffold the iterative, progressive refinement process of creating high-fidelity images (Fig.\,\hyperlink{fig:robin}{1}). The tool takes as input partial sketches where the user denotes each stroke as a blocking or detail stroke. It generates high-fidelity images intended to help the user decide how to adjust the shape and proportion of objects or add additional elements to the composition. 
We demonstrate our system is effective in helping users explore possibilities and progressively refine their vision as they iterate toward a desired final high-fidelity image. 

\vspace{0.5em}
Specifically we make the following contributions:
\vspace{-0.2em}
\begin{enumerate}
    \item A sketch-to-image interface that supports blocking and detail modes of drawing and can generate high-fidelity images from partial control sketches at any stage of completion.
    \item A simple two-pass approach to using blocking strokes as a form of ``loose'' control over spatial composition.
    Our approach uses a ControlNet\,\cite{zhang:2023:controlnet} to generate an image that strictly follows the user's strokes, then adds variation by renoising regions surrounding blocking strokes. 
    Users can control the size of these renoising regions with meaningful (dilation radius) parameters.

    \item A dataset generation scheme that, when used to train a standard ControlNet architecture, allows regions that do not contain strokes to be interpreted as not-yet-specified regions, rather than regions of empty space. This facilitates generation of high-fidelity images from partial sketches that contain only a small number of strokes.
\end{enumerate}

\section{Related Work}
\label{sec:related}

\begin{figure*}
\centering
\includegraphics[width=\textwidth]{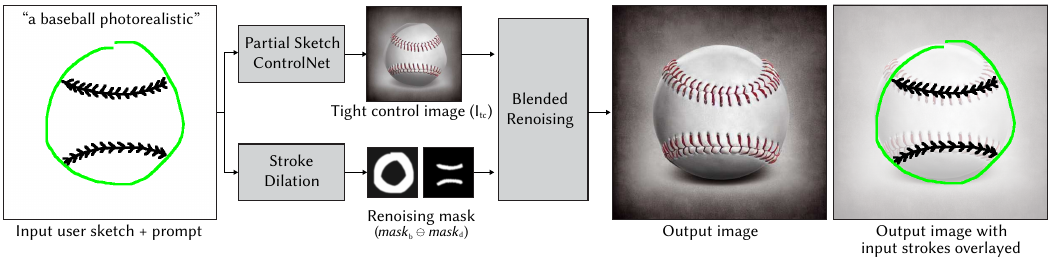}
\vspace{-1em}
\caption{Our algorithm takes as input a text prompt (``a baseball photorealistic''), and a sketch consisting of blocking strokes (green) and detail strokes (black).  
In a first pass it feeds all strokes to a partial-sketch-aware ControlNet to produce an image, denoted as $I_{tc}$, that tightly adheres to all strokes. 
Here the contour of the baseball in $I_{tc}$ is misshapen because the input blocking strokes (green) are not quite circular.
To generate variation in areas surrounding the blocking strokes, our algorithm applies a second diffusion pass we call {\em Blended Renoising}. Based on a renoising mask 
formed by dilating the input strokes, blended renoising generates variation in the area surrounding blocking strokes while preserving areas near detail strokes. The renoised output image corrects the baseball's contour, while the location of the stitching closely follows the user's detailed strokes. 
See Algorithm~\ref{alg:overview} for pseudocode.
}
\label{fig:method}
\vspace{-0.5em}
\end{figure*}

\noindent
{\bf \em Sketch-to-image generation.}

Advances in image generation using GANs\,\cite{isola2017image,zhu2017unpaired} and diffusion models\,\cite{Rombach_2022_CVPR} have led to a variety of efforts aimed at controlling the generation process via user-drawn sketches. GAN-based methods treat this
as a translation problem from the domain of sketches to the domain of high-fidelity 
images\,\cite{chen2018sketchygan,park2019gaugan,liu2021self,liu2020unsupervised}. 
Importantly these methods assume that the training data from the domain of sketches are complete drawings, and therefore do not learn to translate partial sketches into high-fidelity images.
Diffusion-based methods either augment a diffusion model with auxiliary conditioning images representing sketches\,\cite{zhang:2023:controlnet,cheng2023adaptively} or introduce auxiliary loss functions to address the domain gap\,\cite{voynov:2023:sketchdiff}.
However, they either assume access to completed sketches~\cite{zhang:2023:controlnet, voynov:2023:sketchdiff} or operate on limited domains~\cite{cheng2023adaptively}.
Unlike our method, all of these prior sketch-to-image methods treat every stroke in the sketch the same and do not distinguish blocking strokes from detail strokes. Also, because many of them are trained using complete sketches, they treat areas of the sketch that do not contain any strokes as empty space rather than not-yet-specified regions.

\vspace{0.5em}
\noindent
{\bf \em Adding control to text-to-image diffusion.} 
Researchers have explored a variety of methods for controlling text-to-image diffusion models.
One common approach is to develop inference-time modifications of the diffusion process\,\cite{bartal:2023:multidiff,balaji:2023:ediffi,avrahami2023blended,sarukkai2024collage}. A benefit of these approaches is that they are zero-shot and do not require retraining or fine tuning the underlying diffusion model. 
In contrast, ControlNet\,\cite{zhang:2023:controlnet} is a carefully designed fine-tuning approach for adding conditioning in the form of an auxiliary control image (e.g. edge map, pose, depth map, etc.) to a diffusion model. The control image lets users specify the spatial composition of the generated image.  
Researchers have developed a variety of ControlNets offering different types of conditioning controls including stylization images~\cite{chen2023controlstyle}, lighting controls~\cite{controlnet_mysee_2023}, QR codes~\cite{controlnet_qr_pattern_2023}, etc. 
More recent work has proposed ControlNets that offer ``loose control'' over the placement of objects in the output image by treating 
3-D volumes\,\cite{bhat:2023:loosecontrol} as the control signal.
We aim to provide similarly loose controls over spatial composition based on users' blocking strokes.

\vspace{0.5em}
\noindent
{\bf \em Drawing assistants.}
Our goal of providing scaffolding for sketch to image generation is inspired by work on drawing assistants\,\cite{fernquist2011sketch,limpaecher2013real,iarussi2013,matsui2016drawfromdrawings,choi2019sketchhelper,lee2011shadowdraw}. 
These tools aim to help users produce higher quality drawings by providing scaffolds that suggest where to draw strokes, visualize geometric proportions and 3D perspective, and aid in following sketching tutorials. While we also provide scaffolding for users by generating intermediate high-fidelity images that they can use to guide their next strokes, our goal is ultimately to aid users in producing a final high-fidelity image rather than producing the sketch that allowed them to generate it.

\section{Method}
\label{sec:method}

\newcommand{\firstimage}[0]{I_{tc}}
\newcommand{\mask}[0]{mask}

Our system takes as input a partial sketch in which the user has marked each stroke as either a blocking stroke or a detail stroke. The goal is to generate high-fidelity images that serve as scaffolding for the user as they decide how to update their drawing by adding or adjusting strokes.
Thus, these images should:

\begin{itemize}
    \item \textbf{Support blocking.} The images should loosely follow the spatial structure of the blocking strokes, portraying variations in the shape and proportions of objects that the user has blocked out in the input sketch.
    \item \textbf{Respect detail:} The images should tightly match the spatial structure of detail strokes preserving their shape as much as possible.
    \item \textbf{Suggest additions:} The images should contain additional elements (related objects, internal parts, etc.) in regions where there are no strokes in the input sketch. Such additions can help users decide whether other elements should be added to the final composition.
\end{itemize}

\begin{algorithm}[t]
\DontPrintSemicolon
\SetAlgoLined
\SetKwInput{KwInput}{Input}
\SetKwInput{KwOutput}{Output}
\SetKwFunction{FMain}{BlockAndDetail}
\SetKwProg{Fn}{Function}{:}{}
\KwInput{Prompt $p$, \blocking image $b$, \detail image $d$}

$\oplus$: pixelwise $x \lor y$ %

$\ominus$: pixelwise $x \land (\lnot y)$ %

\Fn{\FMain{p, b, d}}{
    $\firstimage \gets \text{PartialSketchControlNet}(p, b \oplus d)$\;
    $\mask_b \gets \text{Dilate}(b, \sigma_b)$\;
    $\mask_d \gets \text{Dilate}(d, \sigma_d)$\;
    $\KwRet ~\text{BlendedRenoising}(p, \firstimage, \mask_b \ominus \mask_d, \sigma)$\;
}
\caption{Our sketch-to-image generation pseudocode}
\label{alg:overview}
\end{algorithm}

Algorithm~\ref{alg:overview}, illustrated in Fig.~\ref{fig:method}, achieves these goals. It takes as input a prompt string $p$, and two binary images $b$ and $d$ that denote the pixels covered by blocking strokes and detail strokes respectively.
The algorithm first uses a ControlNet~\cite{zhang:2023:controlnet} to generate an image ($I_{tc}$) that tightly conforms to \emph{all strokes} in the sketch; both blocking and detail. 
To encourage variation around the blocking strokes, we perform a second diffusion pass we call {\em Blended Renoising}. In this pass we use a renoising mask based on dilating the input strokes to 
preserve regions of $I_{tc}$ around detail strokes, while allowing significant changes to regions surrounding blocking strokes. 
Conceptually, this second pass is initialized by $I_{tc}$, but provides the image generation process ``wiggle room'' to make changes around blocking strokes.

Finally, to better meet our ``suggest additions'' goal, we train the initial ControlNet to generate output from partial sketches that may include only a small number of strokes. In the following sections we first describe how we implement our dilation-based blended renoising to provide loose control near blocking strokes and tight control near detail strokes. We then explain how we train our partial-sketch-aware ControlNet. 

\change{
While our primary goal is to improve user control of the geometry and layout of the composition, users may desire consistency in colors across iterations -- e.g., making sure that the colors in the generated samples remain consistent with the sample selected in the first iteration. We obtain such consistency by adding the latent encoding of the selected sample from a prior  iteration to the initial image seeds for all the images generated in the next iteration. We provide this color consistency as an option that users can decide to turn on or off; all examples in the paper are shown with the off, except for Figs.~\ref{fig:lamp} and ~\ref{fig:rose}---the two figures where we keep color consistency on.}

\subsection{Dilation-based Blended Renoising}
\label{sec:method_blocking}
To generate variation in areas around the blocking strokes $b$ while maintaining a tight match to detail strokes $d$ we renoise $I_{tc}$ using a pair of influence masks $mask_b$ and $mask_d$. We generate $mask_b$ and $mask_d$ by dilating the blocking stroke $b$ and detail strokes $d$ using a dilation radius $\sigma_b$ (typically large) and radius $\sigma_d$ (typically small) respectively.
Subtracting mask $mask_d$ from $mask_b$ yields the desired renoising region.

Given this region mask, we leverage Blended Latent Diffusion~\cite{avrahami2023blended}, an inference-time mechanism for inpainting, and SDEdit~\cite{meng2021sdedit}, which generates variations in image content via injecting noise into existing image content, then running a diffusion denoising. Specifically, we add noise at level $\sigma_{t}$ and perform $40$ denoising steps with the Euler-Ancestral solver to generate variations in these regions. We set $t=0.8$ by default.

\begin{figure}
\centering
\includegraphics[width=3.3in]{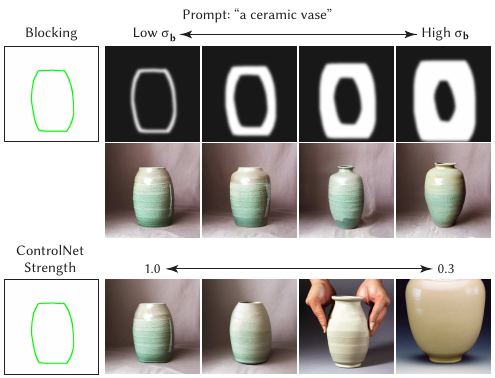}
\caption{Dilation radius is more interpretable as a parameter for loosening spatial control than ControlNet strength.
As the dilation radius of blocking strokes $\sigma_b$ increases, our algorithm maintains the rough shape of the vase while gradually allowing greater variation in the exact shape of its contour. In contrast, varying the ControlNet guidance strength 
without dilating the stokes leads to sudden unpredictable behavior as the strength is decreased. We lose adherence to the strokes as the hands appear in the third image and the vase completely loses its form by the fourth image. 
}
\label{fig:blocking_vs_strength}
\end{figure}

\vishnu{
In prior sketch-to-image tools, adherence to the input sketch is controlled by unpredictable ``guidance strength'' parameters\,\cite{zhang:2023:controlnet,voynov:2023:sketchdiff}.
Instead, our algorithm provides direct control over spatial variability via dilation radius parameters $mask_b$ and $mask_d$. 
} 
These are easily interpretable as setting the amount of spatial variation permitted around blocking strokes (Fig.~\ref{fig:blocking_vs_strength}). 
Note also that our two-pass algorithm can be used with existing sketch-to-image ControlNets in a zero-shot manner without training. It can utilize ControlNet, SDEdit, and Blended Latent Diffusion unmodified as subroutines. However, we show that 
our partial-sketch-aware ControlNet (which does require training) produces better results than the zero-shot approach (Fig~\ref{fig:partial_vs_scribble}).

\subsection{Partial-Sketch-Aware ControlNet}
\label{sec:method_partial}

\begin{figure}[]
\centering
\setlength{\tabcolsep}{1pt} %
\renewcommand{\arraystretch}{1.0} %
\begin{tabular}{ccccc}
Image & 20\% lines & 60\% lines & 80\% lines & 100\% lines \\
\includegraphics[width=0.19\linewidth]{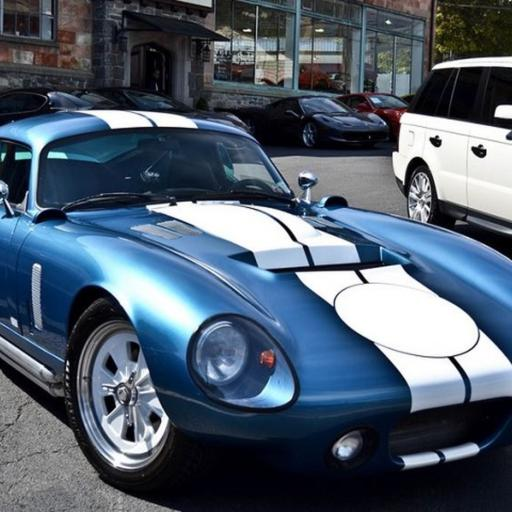} & \includegraphics[width=0.19\linewidth]{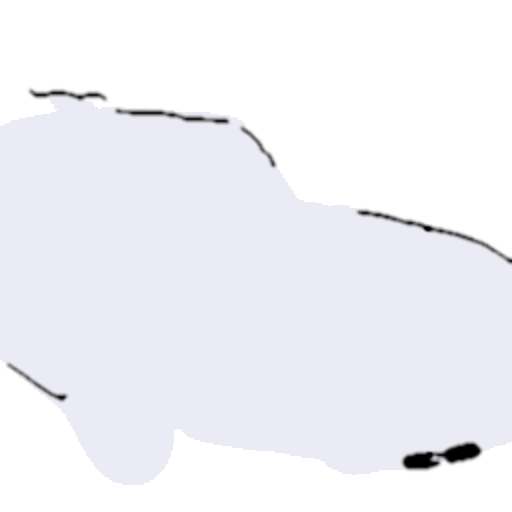} & \includegraphics[width=0.19\linewidth]{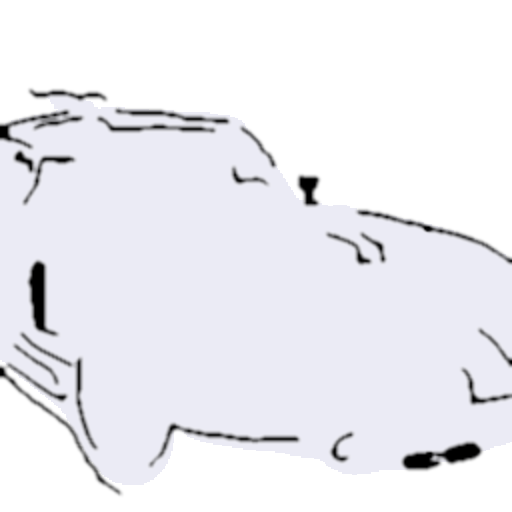} & \includegraphics[width=0.19\linewidth]{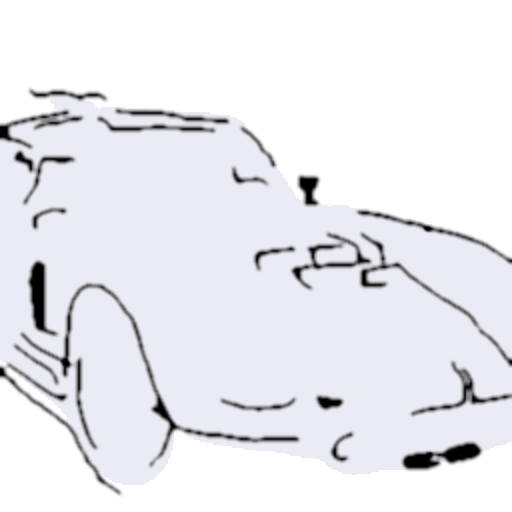}& \includegraphics[width=0.19\linewidth]{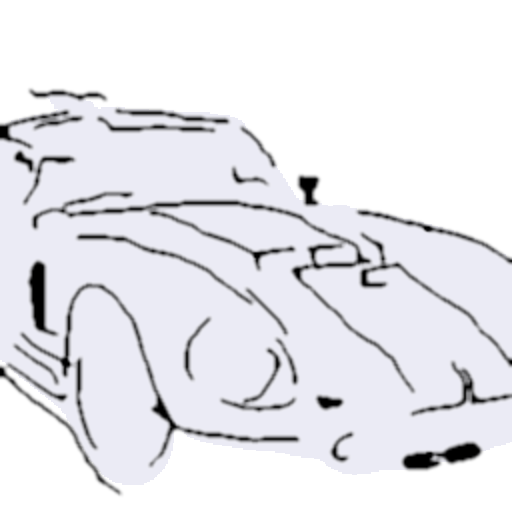} \\
\includegraphics[width=0.19\linewidth]{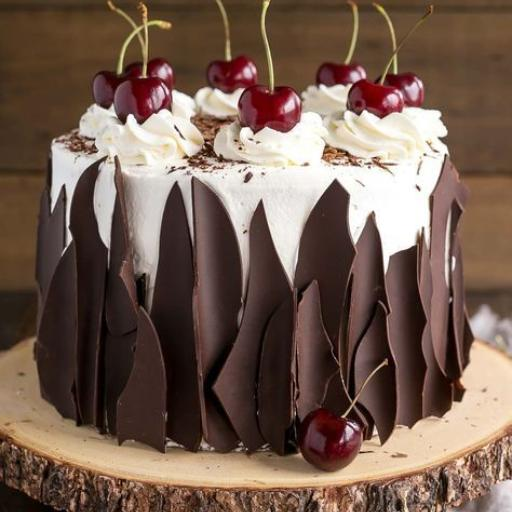} & \includegraphics[width=0.19\linewidth]{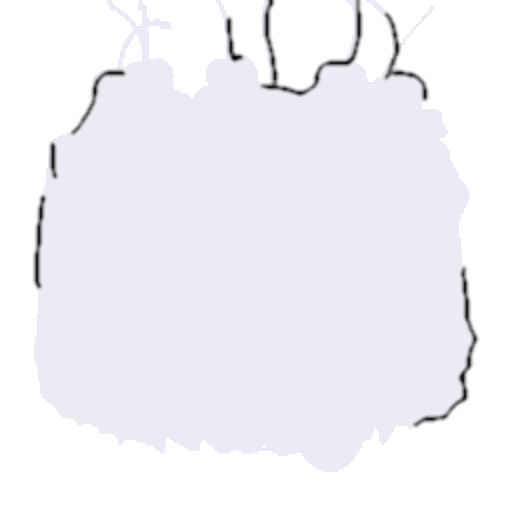} & \includegraphics[width=0.19\linewidth]{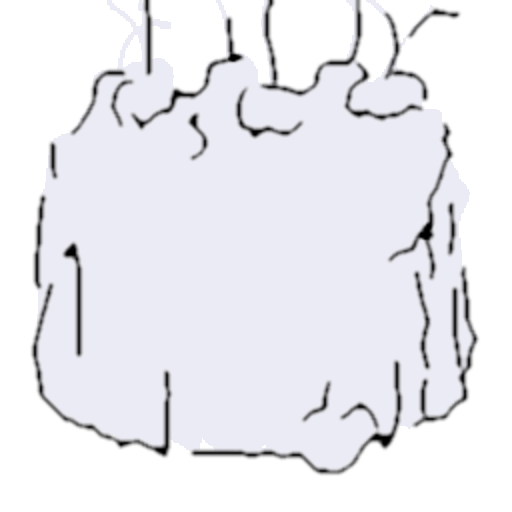} & \includegraphics[width=0.19\linewidth]{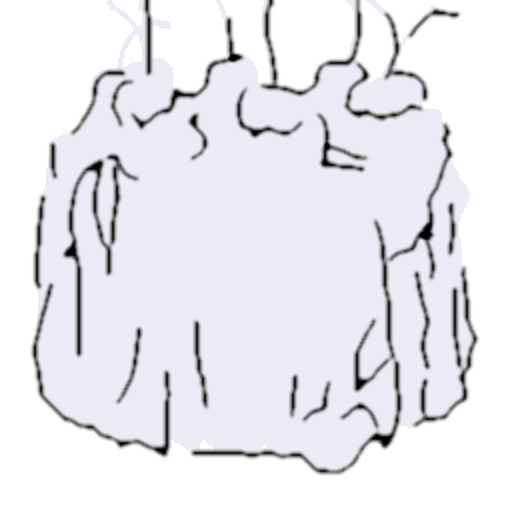} & \includegraphics[width=0.19\linewidth]{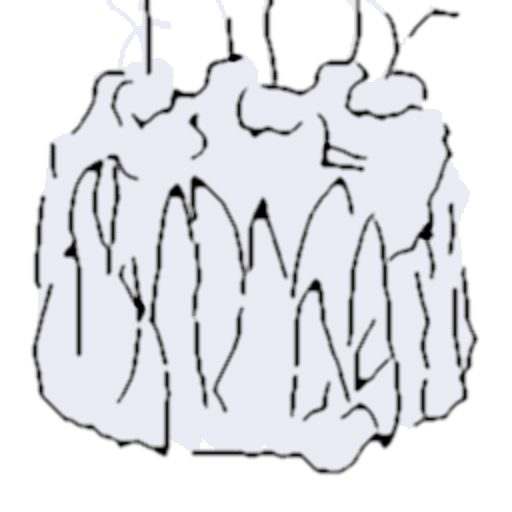} \\
\end{tabular}

\caption{Synthetic data generation: ordering strokes by their distance from the boundary of the foreground object's mask (gray tint) enables the preferential deletion of strokes furthest from the object boundary. 
For the car, the first 20\% of the lines capture a portion of the silhouette. 
This enables us to train a partial-sketch ControlNet that auto-completes object forms and also attempts to fill in details. %
}
\label{fig:synthetic_data}
\end{figure}

Prior sketch-to-image ControlNet models are trained on pairs of high-fidelity images and their edge maps~\cite{zhang:2023:controlnet}. In this configuration, a region of an edge map with no edges corresponds to a region of an original image with no detail. As a result, ControlNet-generated images typically contain objects only in regions where control sketches contain strokes. We seek a solution that does not treat empty space in a sketch as intended whitespace, but as a signal that a user has yet to specify the region's contents. We look to the system to produce images that contain suggestions for how the region could plausibly be filled.
To obtain this behavior, our approach is to produce training data that pairs of high-fidelity images with corresponding \emph{partial sketches} at various levels of completion. 

Given a high-fidelity training image, we extract a pixel edge map using Im2RBTE\,\cite{efthymiadis2022edge}, and then convert the edge map to vectorized strokes using the approach of\,\citet{mo2021general}. 
To produce partial sketches at various levels of completion, we estimate the order in which strokes were likely to have been drawn by an artist, and then delete the last $X\%$ of the strokes (for random $X$). 
Our approach for estimating the stroke drawing order is based 
on the practice artists commonly use of drawing coarse contours first and then progressively adding finer details\,\cite{liu:2024:dynamicsketch,dodson1990keys}. 
Specifically, we assume that training images contain single-object-centric scenes, and extract foreground and background masks from the high-fidelity image using the approach of\,\citet{qin2020u2}.
We then order the vectorized strokes according to their distance from the foreground-background boundary, so that strokes closest to the boundary are ranked first. Fig.~\ref{fig:synthetic_data} shows example partial sketches that result from this process. 

\begin{figure}
\centering
\includegraphics[width=3.3in]{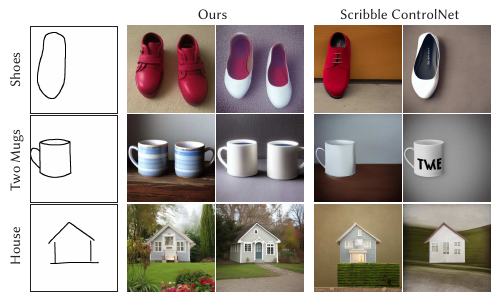}
\vspace{-0.5em}
\caption{Our Partial-Sketch-Aware ControlNet vs Scribble ControlNet\,\cite{zhang:2023:controlnet}. Our Partial-Sketch-Aware ControlNet generates additional elements beyond the drawn strokes: generating multiple shoes instead of a single shoe for `shoes',  generating a second mug for `two mugs', and generating a house with interesting foreground and background details from a simple outline for `house'. The standard Scribble ControlNet fails to generate additional elements and produces minimal details in both the foreground and background of the house.}
\label{fig:partial_vs_scribble}
\vspace{-0.5em}
\end{figure}

Fig~\ref{fig:partial_vs_scribble} illustrates how our partial-sketch-aware ControlNet completes image generations more effectively than standard Scribble ControlNet\,\cite{zhang:2023:controlnet}. Both the `shoes' and `two mugs' sketches demonstrate that our approach is able to generate additional elements when given a sketch of a single object. This capacity to complete scenes also extends to finer-grained detail; our partial-sketch-aware ControlNet produces a variety of plants in the background and details on the windows and doors given the `house' sketch. Scribble ControlNet produces flat images with minimal additional detail especially in the background. By producing additional elements at varying scales in empty regions of the image, our partial-sketch-aware ControlNet provides suggestions that further scaffold the sketching process.

\subsection{Implementation Details}

We train our ControlNet on top of the Stable Diffusion 1.5 backbone~\cite{Rombach_2022_CVPR}. 
The ControlNet is trained via our synthetic sketch generation pipeline on $45000$ images randomly sampled from the LAION-Art dataset \cite{laion5b_2022}.
\vishnu{
At inference time, we use the Stable Diffusion 1.5 backbone in all offline settings with 50 inference steps of the Euler-Ancestral solver.
When testing our interactive tool with artists, we instead leverage the Dreamshaper 7 diffusion backbone\,\cite{Dreamshaper_7}, which is able to generate high-fidelity images with just 8 inference steps of the Euler-Ancestral solver.
}
We run inference on 2 NVidia V100 GPUs. 
\vishnu{
Our complete sketch-to-image algorithm generates images at a speed of $1.0$ seconds per generated image sample when using the 8-step Dreamshaper 7 pipeline, and $2.5$ seconds when using the 50-step Stable Diffusion 1.5 pipeline. 
}

\section{Results}

\begin{figure*}
\centering
\includegraphics[width=7in]{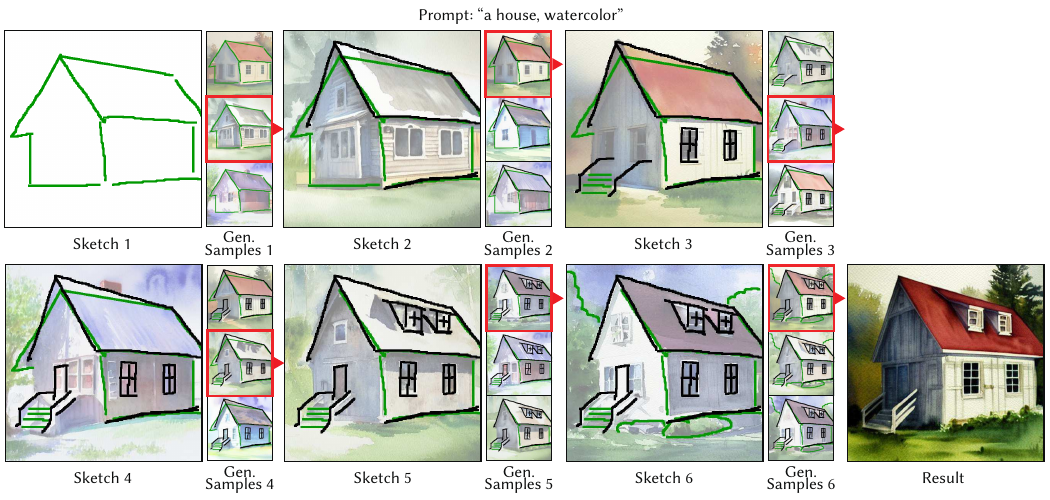}
\vspace{-0.5em}
\caption{Sketch 1: Started by blocking out the general location and orientation of the house. Used the generated image samples 1 for further guidance on perspective and details. Sketch 2: Chose a sample with good perspective, but decided to adjust the shape of the roof with detail strokes to make it taller and angled the base of the house with a detail stroke to match the perspective.
Sketch 3: Took inspiration from the generated samples 2 to add more windows using detail strokes and custom (not found in a sample) stair case railings with blocked in stairs.
Sketch 4: Saw the staircase in generated samples 3, but unsatisfied with the exact shape of it. Adjusted the shape using detail strokes to further refine railings and noticing a lack of entrance in the generated samples from sketch 3 added a doorway.
Sketch 5: Taking inspiration from one of the generated samples 4 to add more details including windows on the roof. Sketch 6:  Observing that generated samples from sketch 5 offer scene completion options such as vegetation near the house, blocked out areas for the vegetation and foliage using blocking strokes. Satisfied with a generated result.  }
\label{fig:house}
\end{figure*}

\begin{figure*}
\centering
\includegraphics[width=7in]{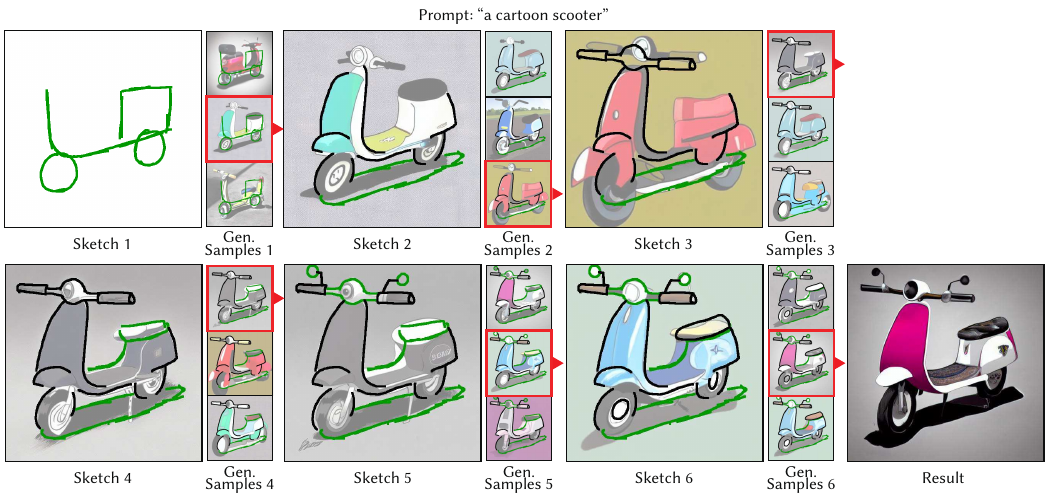}
\vspace{-0.5em}
\caption{Sketch 1: Blocked out the rough shape of a scooter (in cartoon style) to get guidance from the system on size, form and proportions. Sketch 2: The generated samples from sketch 1 revealed better proportions for the wheels and better angles of the main curve marking out the body of the scooter. Traced
one of the samples to lock down the scooter's general proportions and orientation. Left blank spaces for the tool to suggest more variations on the shape of the handles and the seat. Sketch 3: Looked through the generated samples from sketch 2, and traced the shape of a nice handle while moving it to a more desirable location with respect to the scooter body.  Sketch 4: 
Used blocking strokes to roughly define the shape of the seat, taking inspiration from a generated sample from sketch 3. Sketch 5: Added custom rear-view mirrors that were not seen in any of the generated samples from sketch 4. Sketch 6: Added more detail strokes to the tires and to the seat and generated a desired result.}
\vspace{-0.5em}
\label{fig:scooter}
\end{figure*}

\label{sec:results}

\begin{figure*}
\centering
\includegraphics[width=7in]{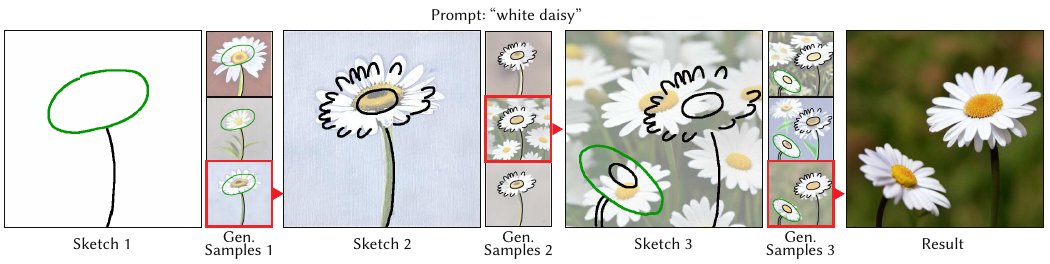}
\caption{Sketch 1: Started by drawing the stem with detail strokes and using blocking strokes to indicate the approximate location of the daisy. Looked for variations in size, orientation and shape in the generated flowers. Sketch 2: Chose a variation
and traced it with detail strokes to lock in the overall shape of the daisy. Sketch 3: Taking inspiration from a generated sample 2 containing multiple daisies decided to have place two daisies in the scene. Moved the strokes for the first daisy to the right 
and added another flower with a mixture of detail and blocking strokes. Chose a final result amongst the generated samples.}
\label{fig:daisy}
\end{figure*}

\begin{figure*}
\centering
\includegraphics[width=7in]{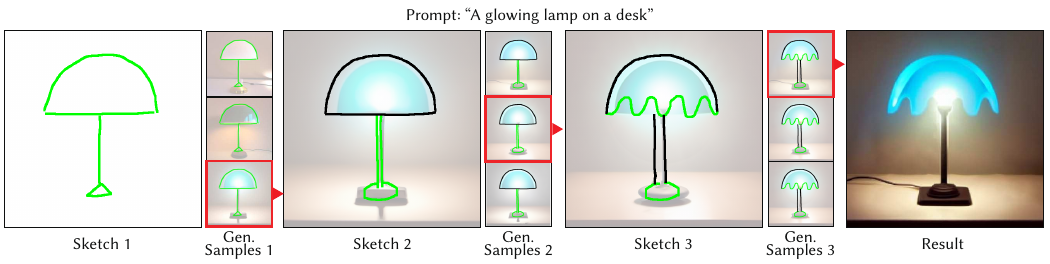}
\vspace{-0.5em}
\caption{
Sketch 1: Blocked out the rough shape of a desk lamp, and generated sample images to check its proportions.
Sketch 2: Observed that the based of the lamp was too small in the generated samples from sketch 1 and replaced blocking strokes at the base. Sketch 3: The generated samples had the desired lamp shape decided to explore other lamp shade designs. 
Added the unusual design element of the wavy lampshade with blocking strokes to obtain variations. Also changed the base to be rounder. The generated images produced a final result.
An example of how users can create image of realistically plausible yet creative object with help from the system. 
\change{
Used the color consistency option for image generation to maintain lamp shade color across iterations.
}
}
\label{fig:lamp}
\end{figure*}

\begin{figure*}
\centering
\includegraphics[width=7in]{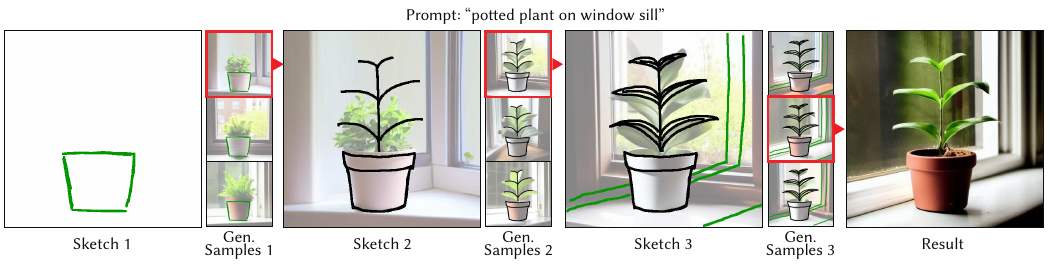}
\caption{Sketch 1: Blocked out the area for the pot. Let the tool  generate variations of plants for inspiration. Sketch 2: Found a desirable shape for the pot from a generated sample and locked down with detail strokes. Having a vague idea for the layout of the plant leaves in mind, added a few detail strokes to orient the leaves. Sketch 3: 
Referenced the generated samples to draw the leaves using detail strokes to express their shape. Also took inspiration for scene completion from the samples and used blocking strokes to indicate the orientation of the background window. Generated a final result.}
\label{fig:pottedplant2}
\end{figure*}

\begin{figure*}
\centering
\includegraphics[width=7in]{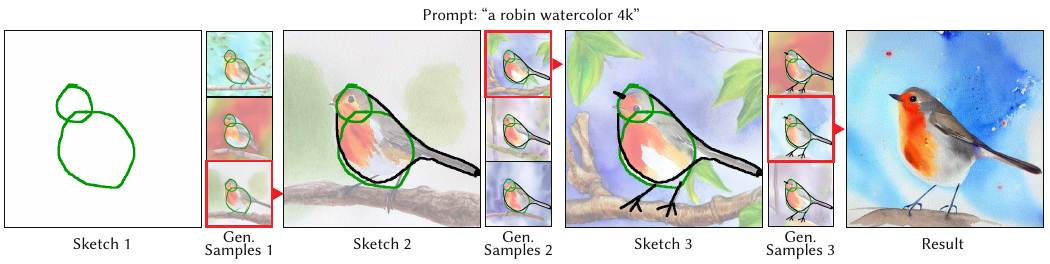}
\caption{Sketch 1: Blocked out the rough shape of the robin’s head and body, to see better proportions in the generated sample. Sketch 2: Liked the tail feather sticking out in one generated sample, so added detail strokes to capture it as well as the bulge in the lower body curving into the head. Sketch 3. Liked the shape of the legs and feet on the branch in a generated sample and added them as detail strokes. All the generated results had beak pointing to the left but wanted it instead to tilt upwards so added detail strokes to point the beak upwards. Generated samples contained the desired robin. }
\label{fig:maneesh_robin}
\end{figure*}

\begin{figure*}
\centering
\includegraphics[width=7in]{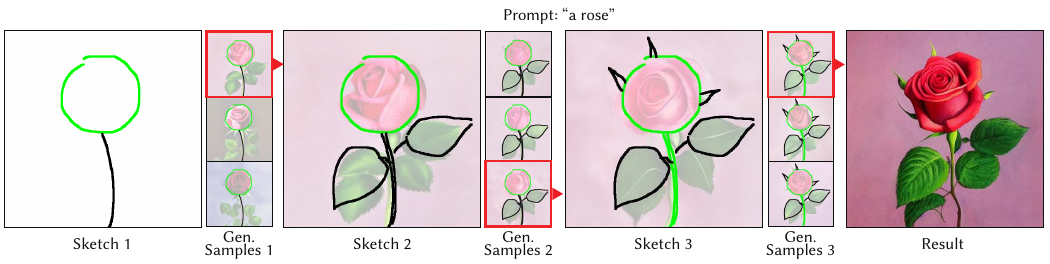}
\vspace{-0.5em}
\caption{
\change{
Sketch 1: Blocked out the rough shape of a rose with detail for the stem, and generated sample images for inspiration.
Sketch 2: Realized that the rose should have some leaves and added these with detail strokes. Sketch 3: Was less certain of the stem so changed these detail strokes to blocking, and added a few thorns to the rose flower. The generated images produced a final result.
Used the color consistency option for image generation to maintain rose and background colors across iterations.
}
}
\label{fig:rose}
\end{figure*}

We present walkthroughs of our system being used to generate a variety of images (Figs.~\hyperref{fig:robin}{1}, ~\ref{fig:house}—~\ref{fig:maneesh_robin}). 
Each figure shows a sequence of sketching steps and the generated image samples for each sketch. 
The supplemental videos also demonstrate how the generated samples scaffold the process.
The walkthroughs illustrate the benefits of blocking, partial-sketch completion, and detailed sketch control for generating images that scaffold the iterative sketch-to-image process. The generated images include plants (Figs.\,\ref{fig:daisy},\,\ref{fig:pottedplant2}), wildlife (Fig.\,\ref{fig:robin},\,\ref{fig:maneesh_robin}), household items (Figs.\,\ref{fig:lamp},\,\ref{fig:pottedplant2}), and architecture (Fig.\,\ref{fig:house}). 
They span different styles including photorealistic (Figs.\,\ref{fig:daisy},\,\ref{fig:lamp}), watercolor (Figs.\,\ref{fig:house},\,\ref{fig:maneesh_robin}), pastel (Fig.\,\ref{fig:robin}) and cartoon (Fig.\,\ref{fig:scooter}). 
\change{Fig.\,\ref{fig:lamp} and Fig.\,\ref{fig:rose} illustrate sketching sequences where the user enabled the color-consistency option across iterations of image generation. }
All walkthroughs include detailed captions on the user's ideation process; here we discuss some trends we observed.

\vspace{0.5em}
\noindent
{\bf \em Blocking.}
Blocking strokes (Section~\ref{sec:method_blocking}) allow variation in form and proportions around the drawn strokes. 
They are primarily used for obtaining suggestions on potential object size, pose and orientation, and then adopting aspects of the generated forms via detail strokes in the next iteration. For instance, blocking strokes inform the proportions of the house in Fig~\ref{fig:house}, the shape of the pot in Fig~\ref{fig:pottedplant2}, the proportions of the scooter in Fig~\ref{fig:scooter}, etc. 
Blocking can also provide a source of inspiration for object details---in Fig~\ref{fig:daisy}, blocking strokes serve as  inspiration for the shape of the center of the flower and its petals.
Blocking strokes can also help users communicate their rough artistic intent to our tool 
in regions where they have no strong idea of the detailed form they want, but only a general notion---in Fig~\ref{fig:house}, the user loosely blocks out foliage around the house.

\vspace{0.5em}
\noindent
{\bf \em Completion.}
Our partial-sketch-aware ControlNet (Section~\ref{sec:method_partial}) generates image content in regions of the sketch that do not contain strokes, that 
can serve as inspiration to the user. 
It enables the generation of a collection of objects when drawing only a single object. In Fig.~\ref{fig:daisy}, the user initially sketches a single daisy but takes inspiration from the generated images to add an additional daisy. 
Sketch completion can also fill in the rest of an object given only some parts, helping the user draw inspiration for the plant given only a sketch of the pot in Fig.~\ref{fig:pottedplant2}, and providing potential forms for the scooter handle and seat in Fig.~\ref{fig:scooter}.
Finally, the variety of image details generated can serve as inspiration for the next strokes added to the sketch, as the user adopts the generated windows both for the house (Fig.~\ref{fig:house}) and potted plant (Fig.~\ref{fig:pottedplant2}), as well as the generated tail for the robin (Fig.~\ref{fig:maneesh_robin}).

\vspace{0.5em}
\noindent
{\bf \em Variations.} 
Images generated by our tool can help users not only by suggesting forms they like, but also by helping users build a sense of what they don't like. 
When none of the generated image samples match the user's desires, additional sketching enables users to introduce novel features.
For instance, upon realizing that the generated samples lack mirrors, the user adds custom rear-view mirrors to the scooter (Fig.~\ref{fig:scooter}), and upon seeing a few different types of foliage, the user draws their own stem and leaves that are completely different for the potted plant (Fig.\ref{fig:pottedplant2}). 
Sketch-based controls enable users to express creative designs that are rare in real-world images and may be unlikely to emerge from text-conditioned diffusion alone. For example, the wavy lamp shade in Fig.~\ref{fig:lamp}. 

\begin{figure*}
\centering
\includegraphics[width=7in]{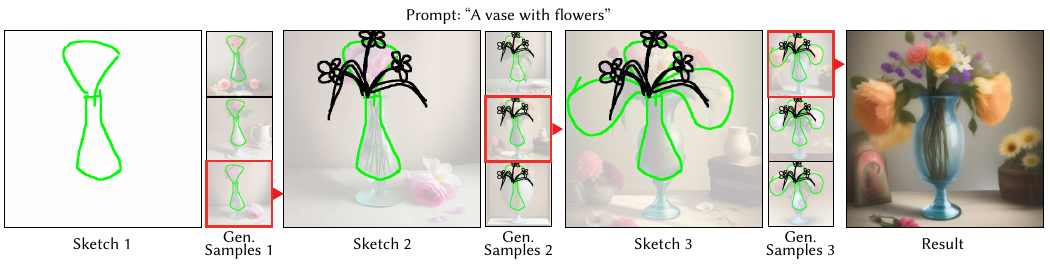}
\caption{ 
\vishnu{
User 1, a vase with flowers. 
Sketch 1: The user started with a unique composition with blocking strokes indicating the vase and flower located on the upper region of the canvas. They expressed their curiosity on the system’s reaction to such input. The system was able to complete the vase with a realistic looking base in all generated samples. 
Sketch 2: The user then added more detail strokes indicating the desired pose of the flowers, hoping to evoke certain feelings with the generated output. The system was able to generate realistic looking flowers and leaves that aligned with the artistic intent. Sketch 3: As a last modification, the user wanted to incorporate larger flowers that draped down from the vase, but they expressed that they were not certain about how to draw flowers with that specific pose, therefore using blocking strokes to allow the system to correct them. As they went through the last batch of generated samples, they found that they actually preferred the system’s generated vase shape, which was slightly thicker than their blocking stroke of the vase. They chose the final image based on the image quality as well as adherence to their desired pose of the flowers.
}
}
\label{fig:yingke_vase}
\end{figure*}

\begin{figure*}
\centering
\includegraphics[width=7in]{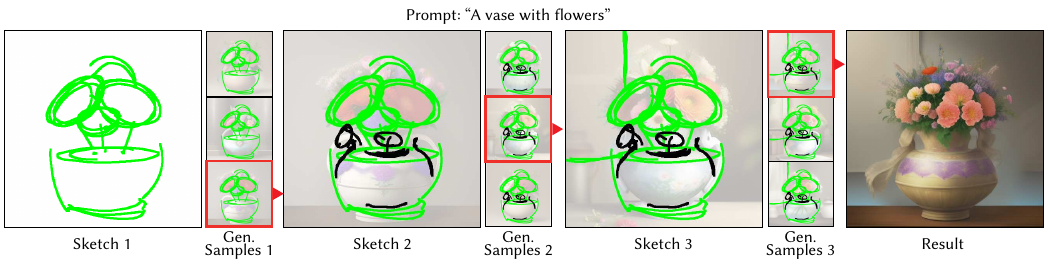}
\caption{ 
\vishnu{
User 2, a vase with flowers.
Sketch 1: They began with a vague idea, sketching the canonical shapes of a vase and flowers with blocking strokes. They browsed through the samples and were pleased with the outputs, which matched the desired composition and proportions of the vase. 
Sketch 2: They picked one particular sample as a reference and added detail strokes to specify the curvature they was looking for in the vase—they wanted the vase “to open up a bit more.” To suggest that the vase’s height be shortened, they added flowers at a specific height, aiming for an even rounder, cuter design of the vase. The system successfully interpreted these cues, generating shorter and rounder vases as desired. 
Sketch 3: They added some blocking strokes to indicate the presence of a rectangular object in the background for compositional purposes, hoping the system would come up with reasonable interpretations. 
In the end, they were satisfied with the final output, as the vase followed the specific shape they wanted and included a rectangular frame in the background that complements the composition of the entire image.
}
}
\label{fig:joon_vase}
\end{figure*}

\begin{figure*}
\centering
\includegraphics[width=7in]{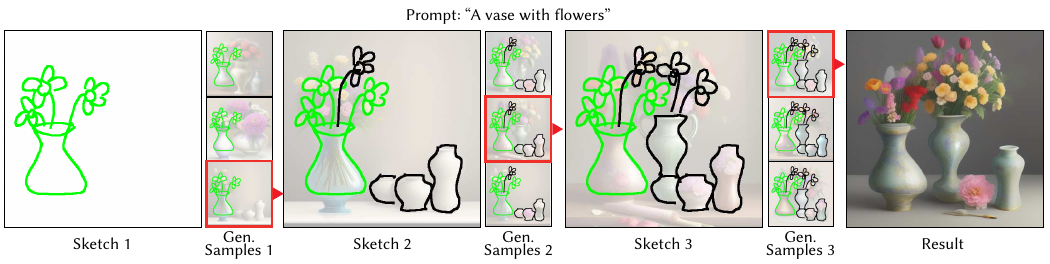}
\caption{ 
\vishnu{
User 3, a vase with flowers. 
Sketch 1:
The user was interested in experimenting with the dilation parameter for the blocking strokes, setting it to a small number (9) to gain more control over the blocking area. They aimed to create a non-traditional composition, positioning the vase of flowers to the very left, and were curious to see what the system would generate on the right. Therefore, they started with blocking strokes that placed the subject on the left. The user was fascinated by the system’s generated samples, which sensibly and nicely positioned related objects, such as other vases and flowers, beside the guided vase. 
Sketch 2:
The user was drawn to a specific generation and used it as a reference to trace over a set of shorter vases using detailed strokes. 
Sketch 3:
They also traced over an additional taller vase that appeared in the next generation step, along with more flowers. 
The final result closely adhered to the specified locations of subjects in the scene. Additionally, the user expressed an interest in having variable strength dilation correlated with each individual stroke. This would allow for more dilation on objects where they needed more assistance, such as the vase on the left, and less dilation on objects they were confident with but still desired some guidance on, such as the flowers.
}
}
\label{fig:oscar_vase}
\end{figure*}

\begin{figure*}
\centering
\includegraphics[width=7in]{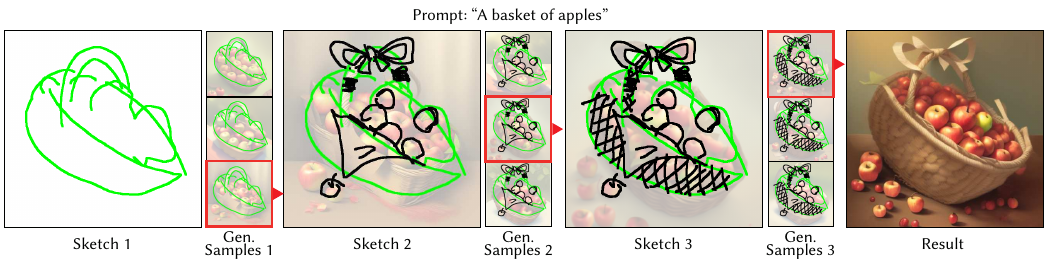}
\caption{ 
\vishnu{
User 1, a basket of apples. 
Sketch 1: The user had a specific design for the basket in mind. They first drew the basket’s overall shape with blocking strokes and were quite happy with the generated samples. 
Sketch 2: They started to add more details using detail strokes; specifically, they wanted to add a bow on top of the basket handle. They were pleasantly surprised by the generated output, which all included a bow on the handle while still following the general shape of the basket. 
Sketch 3: They aimed for the basket to have a certain weaved look, thus adding more detail strokes, which were successfully interpreted by the system. 
In the end, they were satisfied with the final output, which depicted a basket of exactly the shape they wanted, with a bow on the handle, and in the correct style of weaving.
}
}
\label{fig:yingke_apple}
\end{figure*}

\begin{figure*}
\centering
\includegraphics[width=7in]{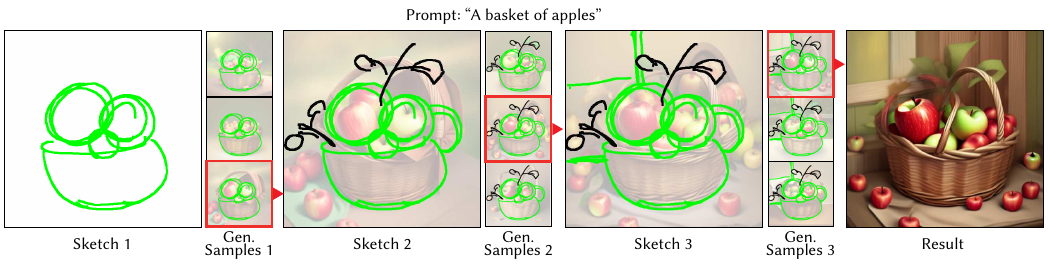}
\caption{ 
\vishnu{
User 2, a basket of apples.
Sketch 1: They put down blocking strokes for the main subject, hoping for more inspiration given by the system. 
Sketch 2: Upon exploring the generated samples, they realized that they wanted to incorporate leaves as part of the composition, putting down detail strokes for two small branches with leaves. A majority of the generated samples matched their intent, while some generated additional basket structures as well. 
Sketch 3: They finished the drawing process by adding some compositional blocking strokes, hoping for generated output to include a window sill, which the tool interpreted successfully. 
They were content with the generated output.
}
}
\label{fig:joon_apple}
\end{figure*}

\begin{figure*}
\centering
\includegraphics[width=7in]{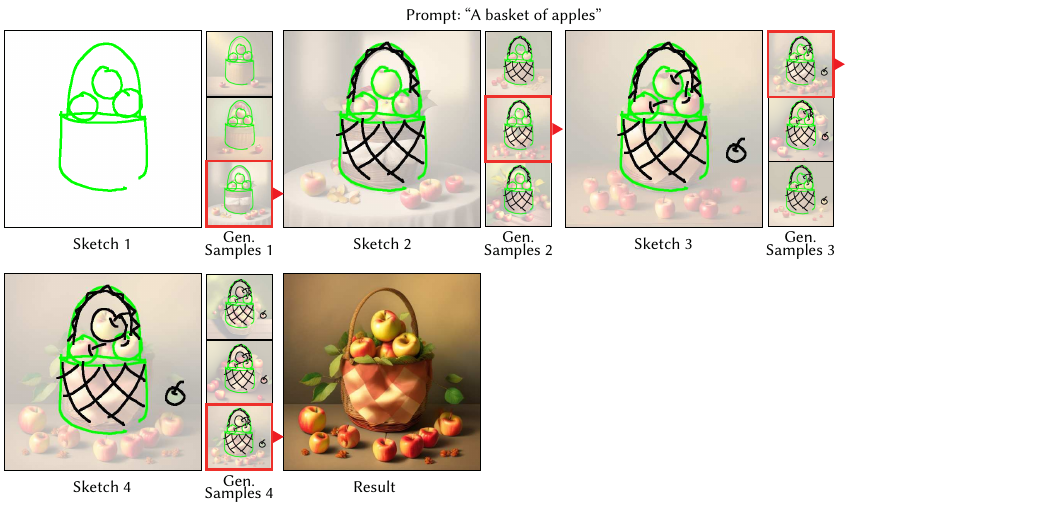}
\caption{ 
\vishnu{
User 3, a basket of apples. 
Sketch 1: 
The user envisioned a particular basket design and began sketching it and some apples using blocking strokes. They expressed satisfaction with the initial creations, particularly praising the realistic appearance. 
Sketch 2: 
They refined the basket and handle with detail strokes, aiming to prompt the system to produce a basket featuring a criss-cross weave pattern. 
The system interpreted the weave pattern as a cloth hanging over the basket, different from the user's intent, but the user was nevertheless pleased by the ideas generated by the system. 
Sketch 3: 
The user further detailed the apples and experimented by changing the blocking stroke of the apple at the highest point to a detailed stroke. They were highly pleased with the final result, which accurately reflected their intended composition, the basket’s pattern, and the arrangement of the apples.
}
}
\label{fig:oscar_apple}
\end{figure*}

\section{User Feedback}
\vishnu{
\change{
We invited three users to generate images using our tool. The users are graduate students aged between 23 and 30, currently enrolled at an American university. These participants brought diverse experiences to the study: one has a professional background in fine arts from their undergraduate studies, another has engaged extensively with art as a hobby, and the third has minimal experience in drawing or design. Despite their varying artistic backgrounds, all participants were familiar with using digital drawing tablets and had experience using generative AI for image creation, albeit without prior use of sketch inputs.
}
}
\vishnu{
The users were given a scripted introduction (approximately 3 minutes) to the system using a slide deck, focusing on explaining the concept of blocking strokes and how they can be useful. All users were able to pick up the interface and start using the tool without further questions, and confusion regarding how to utilize the system was not observed.
Then, they were asked to use the tool to generate images for two prompts: 1) ``a basket of apples'', and 2) ``a vase of flowers''. 
Figs.~\ref{fig:yingke_vase} through~\ref{fig:oscar_apple} illustrate how the three users generated images according to both prompts. User 2 is an experienced artist, while Users 1 and 3 are intermediate artists. 
\change{
We captured all intermediate sketches and images they created during the experiment, and took detailed notes documenting participants’ comments and thought processes. We then analyzed this qualitative data to extract recurring themes and insights into the user experience.
}
}

In early stages of sketching, users found blocking strokes valuable for indicating object form. When users had a specific shape in mind, blocking helped to quickly narrow the space of possibilities--see Figs~\ref{fig:yingke_vase} and~\ref{fig:yingke_apple}, where User 1 had specific notions of the placement of the vase and the shape of the basket respectively. On the other hand, User 2 started by blocking a rough shape of the main object, then searched through the generated images for inspiration--see Figs~\ref{fig:joon_vase} and~\ref{fig:joon_apple}. 

Having blocked out initial object forms, users typically turned to detail strokes to indicate specific forms, often inspired by an image from earlier generations. For instance, User 2 used a generated image as reference for the desired curvature of a vase (Fig~\ref{fig:joon_vase}). However, users still revisited blocking strokes when they wanted to add additional features to the composition--a window sill in Figs~\ref{fig:joon_vase} and ~\ref{fig:joon_apple}, drooping flowers in Fig~\ref{fig:yingke_vase}, etc. 

Overall, users were pleasantly surprised by the quick generation time and were able to maintain momentum throughout the iterative process. The conceptual idea of “Blocking Stroke” was easily understood, and blocking was used both for indicating spatial layout and for specifying rough shapes of objects. Users also were inspired by content from the generated images, often using the generated images to inform object form in subsequent sketches.
Users were consistently satisfied with the quality of generated images throughout the iterative sketching process.

\vspace{0.5em}
\noindent
{\bf \em Numerical evaluations.}
\vishnu{
We conducted a user study comparing our algorithm with Scribble ControlNet~\cite{zhang:2023:controlnet}, a widely used, state-of-the-art approach for sketch-to-image control. 
Given a starting sketch containing both blocking and detail strokes and a text prompt, we generated images with both algorithms and asked users to rank them for overall quality and visible distortion. 
\change{
11 novice viewers evaluated 7 image pairs each. The viewers range in age from 21 to 30. They included five individuals currently enrolled at an American university and six working full-time. Among them, five have professional backgrounds in generative AI. 
}
They preferred the quality of images from our algorithm over Scribble ControlNet for 84\% of the pairs and found our images had less distortion in 81\% of the pairs. 
Detailed information on the image pairs and individual viewer preferences are included in the supplemental material. 
}

\vspace{-0.25em}
\subsection{Limitations}
\vishnu{
While our tools allows users to influence the form and placement of objects through blocking and detail strokes, incorporating a mechanism to specify `negative space' could augment the capabilities of the tool, allowing users to specify where objects should not go. 
One user mentioned that this feature would be useful to their creative process; the feature might offer users more targeted feedback and inspiration. 
}

\section{Conclusion}

Our tool enhances artists' workflows by facilitating sketch-to-image generation that aligns with their vision. The process is iterative, with artists progressively adding and modifying strokes to build and refine the desired image. It mirrors the strengths of diffusion models, which tackle the complex task of image generation through smaller, more manageable steps. By breaking down the image generation into simpler sub-tasks, both the artist and the diffusion model can more effectively address the overall challenge. The critical aspect is to appropriately segment the process into these manageable sub-tasks. Our tool, drawing inspiration from the progressive refinement typical in artistry, aids in this by structuring sketching into stages: blocking, detailing, and completion. As generative models increasingly become integral in artistic endeavors, the decomposition of tasks is crucial in developing efficient human-in-the-loop tools.

\begin{acks}

Support for this project was provided by Meta, Activision, Andreessen Horowitz and the Brown Institute for Media Innovation. 
Thank you to Joon Park, Purvi Goel, James Hong, Sarah Jobalia, and Sofia Wyetzner for their feedback on our tool. 

\end{acks}

\cleardoublepage
\bibliographystyle{ACM-Reference-Format}
\bibliography{blockdetail}

\end{document}